\documentclass[journal]{IEEEtran}
\ifCLASSINFOpdf
\else
\fi

\usepackage{bm}
\usepackage{graphicx}
\usepackage{indentfirst}
\usepackage{epsfig}
\usepackage{amsfonts}
\usepackage{amsmath}
\usepackage{url}
\usepackage{color}
 \usepackage{algorithm}
\usepackage{algpseudocode}
\usepackage{cancel}
\usepackage{graphicx}
\usepackage{mathcomp}
\usepackage{footnote}
\usepackage{cite}
\usepackage{mathtools}
\usepackage{amssymb}
\usepackage{amsmath}
\usepackage{amsthm}
\newtheorem{theorem}{Theorem}

\usepackage{mathrsfs}
\usepackage{amsfonts}
\usepackage{graphicx}
\usepackage[justification=centering]{caption}
\usepackage{multirow}
\usepackage{hhline}
\usepackage{bigstrut}
\usepackage{cases}
\usepackage{verbatim}
\usepackage[utf8x]{inputenc}
\usepackage[T1]{fontenc}
\usepackage{xcolor}
\usepackage{epstopdf}

\begin{document}
\title{Cooperative Planning of Renewable Generations for Interconnected Microgrids}

\author{Hao~Wang,~\IEEEmembership{Member,~IEEE,}
        and~Jianwei~Huang,~\IEEEmembership{Fellow,~IEEE}

\thanks{This work was supported by the Research Grants Council of the Hong Kong Special Administrative Region, China, through the Theme-based Research Scheme under Project No. T23-407/13-N.}
\thanks{H. Wang and J. Huang are with the Network Communications and Economics Lab (NCEL), Department of Information Engineering, The Chinese University of Hong Kong, Shatin, Hong Kong SAR, China, e-mails: \{haowang, jwhuang\}@ie.cuhk.edu.hk.}
}
\maketitle

\begin{abstract}
We study the renewable energy generations in Hong Kong based on realistic meteorological data, and find that different renewable sources exhibit diverse time-varying and location-dependent profiles. To efficiently explore and utilize the diverse renewable energy generations, we propose a theoretical framework for the cooperative planning of renewable generations in a system of interconnected microgrids. The cooperative framework considers the self-interested behaviors of microgrids, and incorporates both their long-term investment costs and short-term operational costs over the planning horizon. Specifically, interconnected microgrids jointly decide where and how much to deploy renewable energy generations, and how to split the associated investment cost. We show that the cooperative framework minimizes the overall system cost. We also design a fair cost sharing method based on Nash bargaining to incentivize cooperative planning, such that all microgrids will benefit from cooperative planning. Using realistic data obtained from the Hong Kong observatory, we validate the cooperative planning framework, and demonstrate that all microgrids benefit through the cooperation, and the overall system cost is reduced by 35.9\% compared to the noncooperative planning benchmark.
\end{abstract}


\IEEEpeerreviewmaketitle

\section*{Nomenclature}

\subsection*{Abbreviations}
\begin{tabular}{l l} 
	CPP &Cooperative planning problem \\ 
	IOP &Investment and operation problem \\
	CSP &Cost sharing problem
\end{tabular}

\subsection*{Indices}
\begin{tabular}{l l} 
	$i$ &Index of interconnected microgrids  \\
	$n$ &Index of users  \\
	$t$ &Index of time slots in the operational horizon\\ 
	$\omega$ &Index of renewable generation scenarios
\end{tabular}

\subsection*{Sets}
\begin{tabular}{l l} 
	$\mathcal{M}$ &Set of interconnected microgrids  \\
	$\mathcal{N}_i$ &Set of users in microgrid $i$ \\
	$\mathcal{H}$ &Planning horizon \\ 
	$\mathcal{T}$ &Operational horizon \\ 
	$\Omega$ &Set of renewable generation scenarios
\end{tabular}

\subsection*{Parameters}
\begin{tabular}{l l} 
	$M$ &Number of microgrids \\
	$D$ &Number of days in the investment horizon \\
	$T$ &Number of time slots in the operational horizon \\
	$R_d$ &Daily discount rate \\
	$\theta$ &Discounted coefficient \\
	$F_i$ &Fixed investment cost \\
	$c_{i}^{s}$ &Investment cost of solar power in microgrid $i$ \\
	$c_{i}^{w}$ &Investment cost of wind power in microgrid $i$ \\
	$\eta_{i}^{s,\omega,t}$ &Solar power profile of microgrid $i$ in $t$ and $\omega$ \\
	$\eta_{i}^{w,\omega,t}$ &Wind power profile of microgrid $i$ in $t$ and $\omega$ \\
	$Q_{i}^{\max}$ &Maximum power procurement of microgrid $i$ \\
	$S_{i}^{\min}$ &Minimum energy storage level in microgrid $i$ \\
    $S_{i}^{\max}$ &Maximum energy storage level in microgrid $i$ \\
    $E_i$ &Capacity of energy storage in microgrid $i$ \\
    $DoD_{i}$ &Maximum depth-of-discharge in microgrid $i$ \\   
	$r_{i}^{\max}$ &Energy storage charge limit in microgrid $i$ \\ 
	$d_{i}^{\max}$ &Energy storage discharge limit in microgrid $i$ \\
	$\eta_{i}^{r},\eta_{i}^{d}$ &Charge, discharge efficiencies in microgrid $i$\\
	$\eta_{i,j}$ &Distribution efficiency between microgrids $i$ and $j$ \\
	$b_{i}^{t}$ &Aggregate inelastic load of microgrid $i$ in $t$ \\
	$L_{n}$ &Total elastic load of user $n$ \\
	$l_{n}^{t,\min}$ &Minimum elastic load of user $n$ in $t$ \\
	$l_{n}^{t,\max}$ &Maximum elastic load of user $n$ in $t$ \\
	$y_{n}^{t}$ &Original load of user $n$ in $t$ \\
	$p^{t}$ &Price of grid power in time slot $t$ \\
	$\alpha_{i}$ &Cost coefficient of storage operation in microgrid $i$ \\
	$\beta_{n}$ &Discomfort cost coefficient of user $n$ \\
    $\pi_{\omega}$ &Realization probability of renewable scenario $\omega$
\end{tabular}

\subsection*{Variables}
\begin{tabular}{l l} 
	$z_i$ &Renewable generation installation in microgrid $i$ \\
	$G_{i}^{s}$ &Capacity of solar power in microgrid $i$ \\
	$G_{i}^{w}$ &Capacity of wind power in microgrid $i$ \\
    $q_{i}^{\omega,t}$ &Grid power procurement of microgrid $i$ in $t$ and $\omega$ \\
    $g_{i}^{\omega,t}$ &Renewable power supply of microgrid $i$ in $t$ and $\omega$ \\
    $e_{i,j}^{\omega,t}$ &Renewable power from microgrid $j$ to $i$ in $t$ and $\omega$\\
	$s_{i}^{\omega,t}$ &Energy storage level of microgrid $i$ in $t$ and $\omega$ \\ 
	$r_{i}^{\omega,t}$ &Energy storage charge of microgrid $i$ in $t$ and $\omega$ \\ 
	$d_{i}^{\omega,t}$ &Energy storage discharge of microgrid $i$ in $t$ and $\omega$ \\ 
    $x_{n}^{\omega,t}$ &Elastic load schedule of user $n$ in $t$ and $\omega$ \\
    $v_{i}$ &Investment cost shared by microgrid $i$ 
\end{tabular}

\section{Introduction}
Recent years have witnessed a significant increase of the share of renewable energy in the overall energy generation profile worldwide. However, the time-varying and intermittent nature of renewable energy makes its integration into the main grid very challenging. Microgrid \cite{mgs}, as one of the key smart grid technologies, can help with the integration and management of distributed renewable energy generations. To prepare for possible independent operation from the main grid, a microgrid often needs to have a total generation capacity that exceeds its critical local load, often in the form of renewable energy investment. On the other hand, renewable energy installation can be expensive, hence underutilization of the installed renewable capacity would be a significant economic loss.

The above observation motivates the recent studies on power grid planning and integration of renewable energy. Specifically, studies in \cite{invest1} and \cite{invest2} examined renewable energy investment strategies through empirical (or numerical) approaches, without considering the tradeoff between investment and operation. Cai \emph{et al.} \cite{invest3} formulated the generation capacity optimization problem with inelastic demands, without considering consumers' demand responses. Yang and Nehorai \cite{addinvest1} studied a planning problem for energy storage and generators in a microgrid, and formulated a joint optimization problem to minimize the total investment and operational cost. Jin \emph{et al.} \cite{invest4} studied the impact of demand response on the thermal generation investment. The studies in \cite{invest1,invest2,invest3,addinvest1,invest4} all focused on capacity investment problems from a single microgrid operator or planner's perspective. Renewable energy generations and load profiles vary in different geographical locations and at different time periods of a day. Baeyens \emph{et al.} \cite{diverse} showed that aggregating diverse renewable resources from geographically distributed areas can substantially reduce the generation variability. This has motivated research towards planning and operation of distributed renewable sources in \cite{review1,review2,review3}.

Planning of renewable sources in microgrids requires comprehensive evaluation of both the initial investment and its subsequent impact on the operation. This requires us to jointly consider the system optimization at two different time scales: the long-term planning horizon and the short-term operational horizon. Moreover, different from the traditional power grid operation, microgrids are often designed to be self-operated, and hence have their own local interests. This brings challenges to the cooperative planning and operation of multiple microgrids. Therefore, an incentive mechanism is needed to encourage cooperation among independent microgrids in generation capacity planning. 

In our previous work \cite{ourpaper1}, we studied the renewable generation planning in a single microgrid. In \cite{ourpaper2}, we studied the interaction of multiple microgrids in a distribution network, assuming that the investments are given in each microgrid. In this paper, we aim to study the more challenging planning problem of multiple interconnected microgrids, to explore diverse renewable resources at different locations. In particular, interconnected microgrids cooperatively decide the optimal renewable generation capacities for long time period (say several years), and manage power supplies, energy storage units, and demand responses, and energy trading over many short time periods (such as on a daily basis). Compared with our previous work \cite{ourpaper1,ourpaper2}, the cooperative planning problem is more challenging in the following aspects: (i) each microgrid's decisions involve two coupling periods: planning and operation, and each microgrid’s decisions on capacity planning and power scheduling are also coupled with other microgrids’ decisions; (ii) renewable generation profiles exhibit diversities across locations and technology types. We seek to understand and take advantage of the diversity, and develop a holistic theoretic framework for data analysis and optimal decision.

The main contributions of this paper are as follows.
\begin{itemize}
	\item \textit{Meteorological data analytics}: Based on meteorological data acquired from the Hong Kong Observatory, we study the potentials of solar and wind energy generations and their correlations across different locations of Hong Kong. The results show diverse profiles of renewable energy generations in terms of technologies and locations, which motivate us to study the cooperative planning of renewable energy generations.
	
	\item \textit{Cooperative planning framework}: We develop a theoretical framework that leads the optimal investment strategies in deploying different types of renewable generations across different locations. We model the planning problem as a cooperative game, in which microgrids cooperatively decide the renewable energy investment levels at all microgrids and the corresponding cost sharing based on the Nash bargaining framework. 
	
	\item \textit{Numerical case studies based on realistic data}: We conduct numerical case studies based on realistic meteorological data of Hong Kong, and compute the optimal planning of renewable generations and fair cost sharing. We show that our proposed cooperative planning framework can reduce 35.9\% of the overall cost compared with the noncooperative approach.
\end{itemize}

The rest of this paper is organized as follows. We analyze the renewable energy generations of Hong Kong in Section II, and formulate the interconnected-microgrids system in Section III. We propose the cooperative planning problem and design the cost sharing scheme in Section IV. Numerical studies are presented in Section V, and we conclude in Section VI.

\section{Renewable Generations in Hong Kong}
To study the renewable power generations in Hong Kong, we acquire meteorological data from the Hong Kong Observatory. The data include the hourly solar radiation data measured at King's Park (KP), and hourly wind speeds measured at seven different locations of Hong Kong: KP, Tai Mei Tuk (TMT), Sha Tin (SHA), Sai Kung (SKG), Tate's Cairn (TC), Tai Po Kau (TPK), and Waglan Island (WGL). Since Hong Kong is a relatively small area, we assume that the solar radiation is the same across the entire Hong Kong and can be represented by the solar radiation at KP.

\subsection{Renewable Energy Potential and Correlation}
We first study the renewable generations from solar and wind at seven locations of Hong Kong, and then analyze the potentials and correlations of different technologies (solar and wind energy) across different locations.

Solar power and wind power generations highly depend on the solar radiation level and wind speed, respectively. We denote the hourly solar radiation as $I^{d,t}$ and hourly wind speed as $V^{d,t}$, where $t \in \{1,2,...,T \}$ is the hour index, $T=24$, and $d \in \{1,2,...,365 \}$ is the day index within an entire year. The hourly solar radiation is measured in $Wm^{-2}$, corresponding to the solar radiation energy received on a unit surface area on earth. The hourly wind speed is measured in $m/s$, which corresponds to the distance traveled per unit of time.\footnote{For presentation clarity, we omit the location index for the solar radiation and wind speed in Section II.}

The power generated from a solar module can be calculated using the following formula \cite{model1}:
\begin{align}
p_{s}^{d,t}=A_{m}\eta_{m}P_{f}\eta_{c} I^{d,t}, \label{solarpower}
\end{align}
where $A_{m}$ is the solar cell array area, $\eta_{m}$ is the module
reference efficiency, $P_{f}$ is the packing factor, and $\eta_{c}$
is the power conditioning efficiency.

Regarding the wind speed, we denote $V_{ci}$ and $V_{co}$ as the cut-in and cut-out wind speed. The wind power will be zero when the speed is less than $V_{ci}$ or above $V_{co}$. The latter case is due to the protection of wind turbine under a very high wind speed. When the wind speed is between $V_{ci}$ and $V_{co}$, the wind power output \cite{model2} can be modeled as
\begin{align}
p_{w}^{d,t}=\frac{1}{2}\rho C_{p}A(V^{d,t})^{3}, \label{windpower}
\end{align}
where $\rho$ is the density of the air, $C_{p}$ is a coefficient related to the performance of the wind turbine, and $A$ is the swept area of the turbine blades.

\begin{figure}[h]
	\centering
	\includegraphics[width=8.0cm]{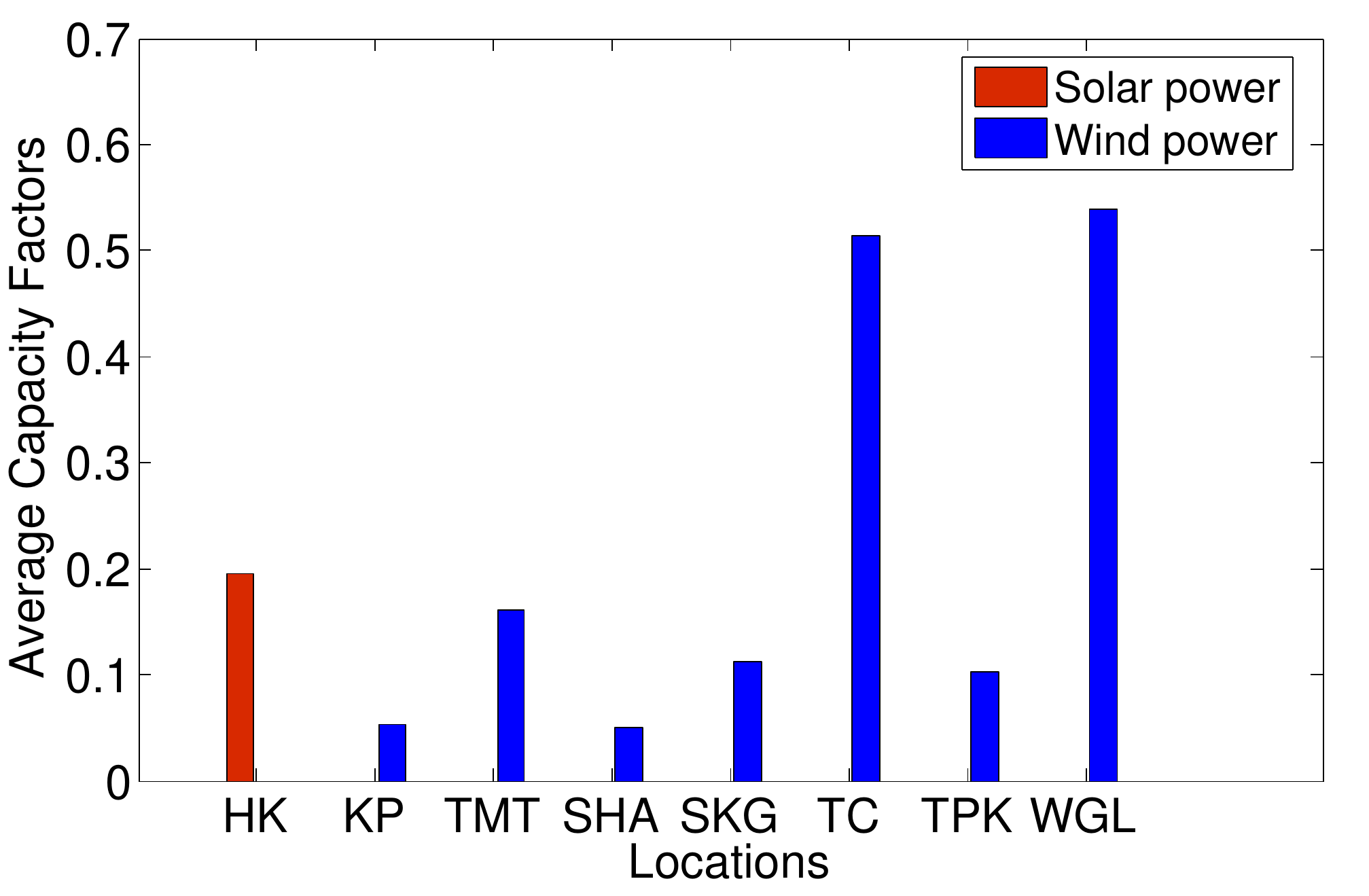}
	\caption{\label{fig-average} Average capacity factors at different locations. Here HK means for the entire Hong Kong.}
\end{figure}

To study the potential of renewable generation, we calculate the average capacity factor of solar power and wind power at different locations. Specifically, the capacity factor is the ratio of the output power to the capacity (maximum possible output power) \cite{correlation}. We plot the average capacity factor of both solar and wind power at seven locations in Fig. \ref{fig-average}. We can see that the average capacity factor of solar power is higher than most of the average capacity factors of wind power, except for TC and WGL, which suggests solar power may be a better choice in location KP, TMT, SHA, SKG and TPK in terms of the generation potential. However, average capacity factors of wind power in TC and WGL are quite high, which suggests high investment return of wind power in TC and WGL.

\begin{figure}[h]
	\centering
	\includegraphics[width=7.8cm]{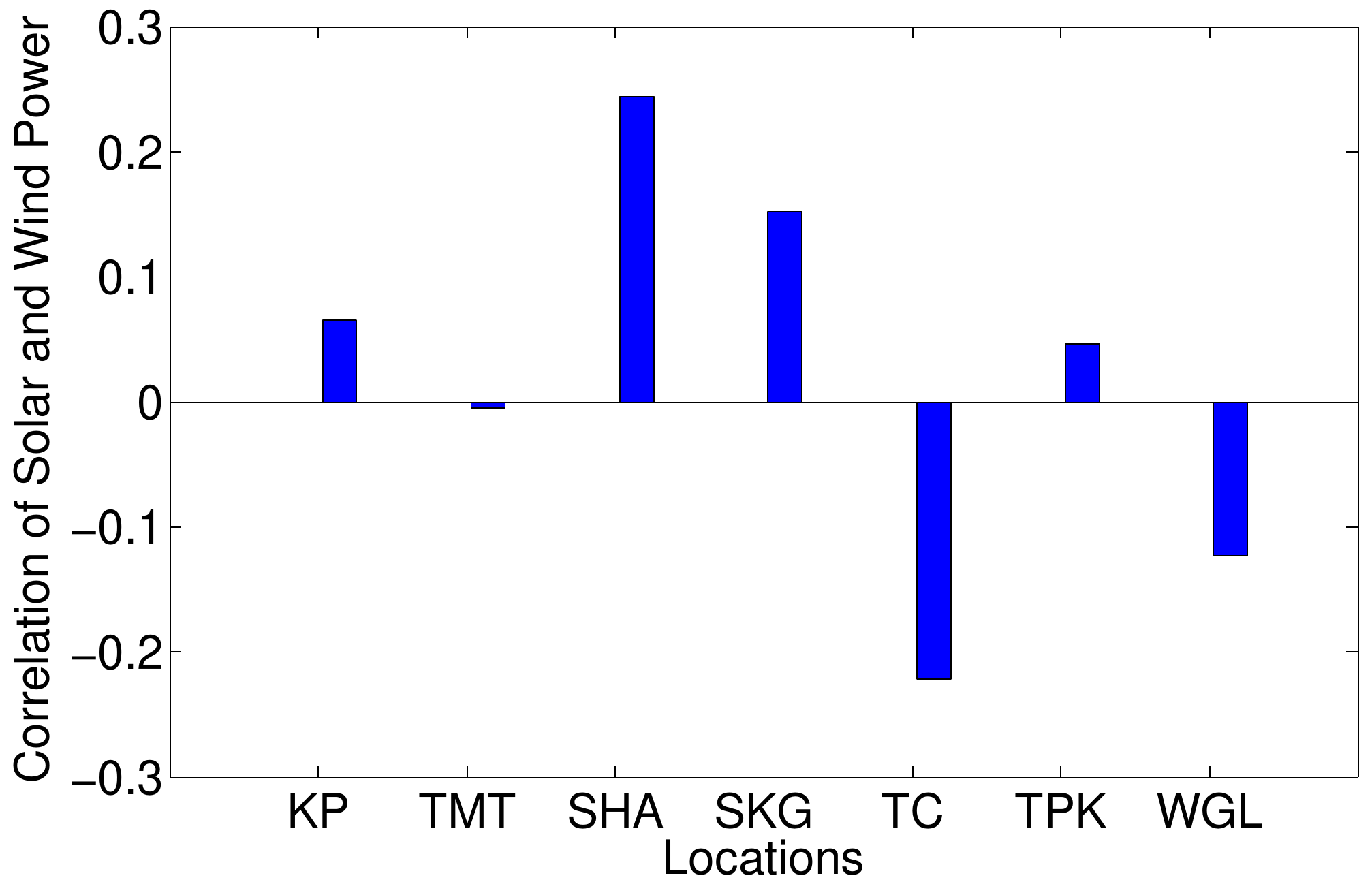}
	\caption{\label{fig-swcorrelation} Solar and wind power correlation at different locations of Hong Kong.}
\end{figure}

Furthermore, we study the statistical correlation between the hourly solar and wind power productions over one year, and calculate the sample correlation coefficient \cite{correlation} as
\begin{align*}
\rho_{X,Y} = \frac{\sum_{k} \left(X(k)-\bar{X} \right) \left(Y(k)-\bar{Y} \right)} {\sqrt{\sum_{k} \left(X(k)-\bar{X} \right)^{2}}  \sqrt{\sum_{k} \left(Y(k)-\bar{Y} \right)^{2}} },
\end{align*}
where $X$ and $Y$ are data series with $k=1,...,K$ terms, $\bar{X}$ and $\bar{Y}$ are the mean values of $X$ and $Y$, respectively, and $\rho_{X,Y}$ measures the correlation coefficient between $X$ and $Y$. We substitute the one-year hourly solar power production into $X$, and the one-year hourly wind power production into $Y$, and calculate the correlation between solar power and wind power of each location, shown in Fig. \ref{fig-swcorrelation}. We find that the wind powers at four locations (KP, TPK, SHA, SKG) have positive correlations with solar power, while the correlations are negative at the other three locations (TMT, TC, WGL). Solar power and wind power complement each other, especially at locations with negative correlations. We will show that the optimal planning mixes negatively correlated renewable generations later in Section V.

Similarly, we calculate the statistical correlation of each pair of wind powers across all the locations, and summarize the results in Table \ref{table-correlation}. We see that all the correlation coefficients are positive. Therefore, wind power at different locations may substitute each other.
\begin{table}[!htb]
	\caption{Correlation of wind power across different locations of Hong Kong}
	\label{table-correlation}
	\centering
	\begin{tabular}{c| c| c| c| c| c| c| c}
		\hline \hline
		Correlation & KP & TMT & SHA & SKG & TC & TPK & WGL	 \\
		\hline
		KP & 1.00 &	0.56 & 0.38 & 0.46 & 0.49 &	0.63 & 0.47 \\
		TMT & 0.56 & 1.00 &	0.37 & 0.52 & 0.49 & 0.73 &	0.51 \\
		SHA & 0.38 & 0.37 & 1.00 & 0.36 & 0.26 & 0.23 & 0.31  \\
		SKG  & 0.46 & 0.52 & 0.36 & 1.00 & 0.33 & 0.50 & 0.37 \\
		TC  & 0.49 & 0.49 & 0.26 & 0.33 & 1.00 & 0.43 & 0.73 \\
		TPK & 0.63 & 0.73 & 0.23 & 0.50 & 0.43 & 1.00 &	0.41 \\
		WGL & 0.47 & 0.51 & 0.31 & 0.37 & 0.73 & 0.41 & 1.00  \\	
		\hline \hline
	\end{tabular}
\end{table}

\begin{figure*}[t]
	\begin{center}
		\begin{minipage}[c]{0.95\textwidth}
			\includegraphics[width=18cm]{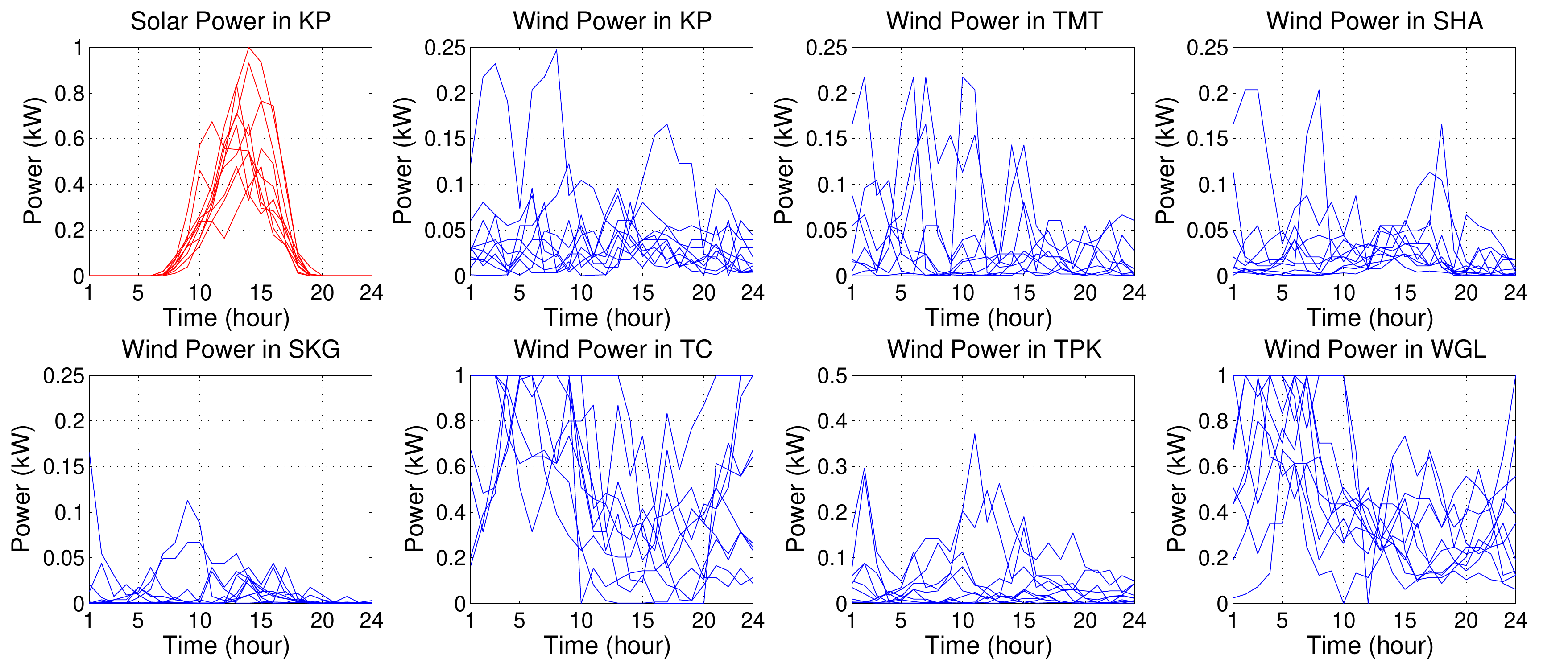}
			\caption{\label{fig-scenario} Renewable energy scenarios (including 10 daily productions of solar and wind power across seven locations).}
		\end{minipage}
	\end{center}
\end{figure*}

The renewable generation profiles exhibit a remarkable diversity, which motivates us to study the cooperative planning of renewable energy across technologies and locations. For example, the users at those locations with low potential of renewable energy have more incentive to cooperate with others who have high renewable energy output, especially for wind power. Solar and wind power generations also show locational patterns. For those locations with negative correlation between solar and wind power generations, it is easier to obtain relatively stable renewable energy generation when investing both technologies; while for other locations one needs to further reply on energy storage and demand response program to achieve relatively stable renewable energy generation with significant more costs. Wind power correlations are positive, and thus a high wind power production at one location can provide supply for several locations.

\subsection{Renewable Energy Scenarios}
For the purpose of later optimization formulations, we model the renewable generations as a set of daily realizations of hourly solar and wind power productions \cite{ourdata}. Each realization of daily power production is called a \emph{scenario}\footnote{The typical practice of power market is based on hourly power scheduling and billing, and this is the reason that we generate 24-hour power production as one scenario.}, denoted by $\omega$. Specifically, each renewable energy scenario is represented by the joint 24-hour solar and wind power productions of all seven candidate locations. Based on one-year data, we have the total number of original scenarios $\widetilde{S} = 365$. The corresponding realization probability of each original scenario is given as $\widetilde{\pi}_{\omega}=\frac{1}{365}$, $\omega = 1,...,\widetilde{S}$.

Due to the large number of original scenarios, the computation later on can be intractable. Thus, it is very useful in practice to approximate the original large set of scenarios with a much smaller subset that can well represent the original scenario set. We use the scenario reduction algorithms \cite{scenario,ourpaper1} to determine a scenario subset and assign new probabilities to the preserved scenarios, such that the corresponding reduced probability measure is the closest to the original measure in terms of the probability distance between the two probability measures. After reduction, the total number of reserved scenarios is denoted as $S$, and the scenario set is $\Omega = \{ 1,...,S \}$. The new realization probability of each scenario is denoted as $\pi_{\omega}$, and $\sum_{\omega \in \Omega} \pi_{\omega}=1$. 

For the purpose of illustration in this paper, we set the number of preserved scenarios as 10, and generate selected scenarios for the solar power generation and wind power generations, depicted in Fig. \ref{fig-scenario}. The actual number of scenarios $S$ depends on the tradeoff between performance and complexity in practice. 

Fig. \ref{fig-scenario} shows the renewable generations (both solar power and wind power) per kW capacity of investment, respectively. We see that the solar power has a peak at noontime, while wind power productions show dramatic locational differences. Wind power at WGL is often adequate during night time, while wind power at TPK reaches a higher output level during day time. In addition, wind power at TC and WGL has a higher average output than that at other locations, which implies that TC and WGL have more potentials for wind power production. The diverse renewable generations motivate us to study the cooperative planning of renewable generations.

\section{System Model}
\begin{figure}[h]
	\centering
	\includegraphics[width=7.5cm]{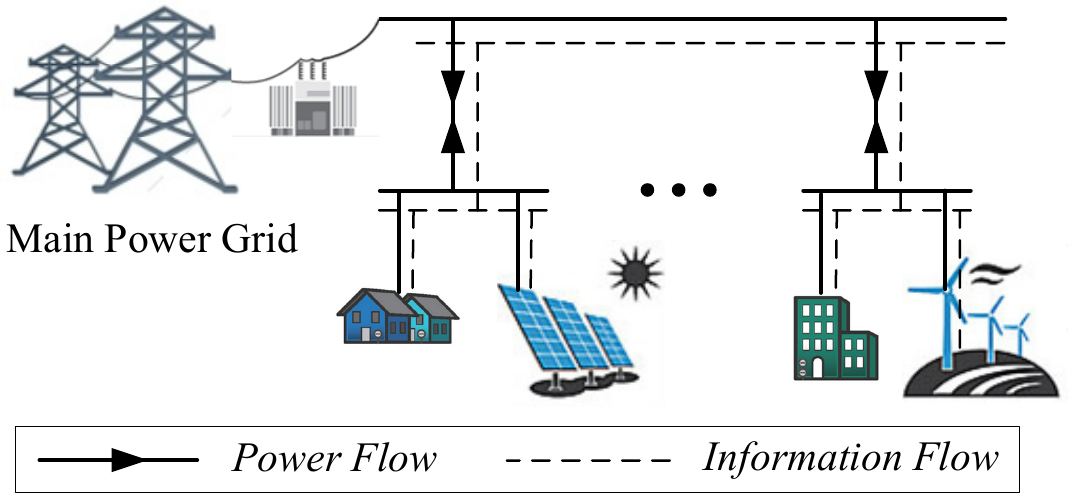}
	\caption{\label{fig-system} System architecture.}
\end{figure}

Consider a distribution network including a set $\mathcal{M}=\{1,...,M \}$ of interconnected microgrids, all of which are connected to the main power grid as well as with each other through the distribution bus, as shown in Fig. \ref{fig-system}. Each microgrid $i \in \mathcal{M}$ owns some energy storage, and has implemented the demand response program. Each microgrid is capable of investing both solar and wind renewable energy generations, and the actual investment amounts are the variables to be optimized. We further assume that each microgrid has space to deploy renewable energy at its own location.

To explore the diversities of renewable energy generation potentials at different locations, the interconnected microgrids jointly plan the renewable generations. Each microgrid needs to consider the interactions with other interconnected microgrids and the impact of the long-term investment on its future short-term daily local power scheduling and the operational cost. In particular, renewable generation investment determines the availability of renewable power outputs in the next few years\footnote{We consider 20 years as the planning horizon in the later case study.}, and thus affects the future daily operational cost. On the other hand, the accumulative operational costs can be substantial over a long period of time, and should be considered when planning renewable generation investment. The interactions among microgrids will enable the exploitation of diversity across locations, and hence improve the overall system efficiency through a proper incentive mechanism.

\subsection{Renewable Generation Investment}
We consider an investment horizon $\mathcal{H}=\{1,...,D\}$ of $D$ days, and let $z_i=\{0,1 \}$ denote the long-term investment decision of microgrid $i$. Usually, renewable generation facilities (\emph{e.g.}, photovoltaic panels and wind turbines) occupy large space, which leads to a significant fixed cost of installation (besides the additional cost depending on the capacity). To account for the locational difference, we denote $F_i$ as the fixed investment cost in microgrid $i$.

We assume that each microgrid has two candidate renewable sources: solar and wind. If microgrid $i$ decides to install renewable generation, \emph{i.e.}, $z_i=1$, it needs to determine the capacities of solar power $G_i^{s} \in [0,G_i^{s,\max}]$ and wind power $G_i^{w} \in [0,G_i^{w,\max}]$, both measured in $kW$, where $G_i^{s,\max}$ and $G_i^{w,\max}$ are the maximum capacities allowed for solar power and wind power deployment at location $i$. The capital investment cost for microgrid $i$ is
\begin{align*}
& C^{I}_{i}(z_i,G_i^{s},G_i^{w}) = z_i (F_i + c_i^{s} G_i^{s} + c_i^{w} G_i^{w}),
\end{align*}
where $c_i^{s}$ and $c_i^{w}$ denote the investment cost of solar and wind generation per $kW$ in microgrid $i$. We assume that the investment cost covers all expenditures, \emph{e.g.}, installation and maintenance of photovoltaic panel for solar energy, turbine for wind energy, controllers, inverters, and cables. These invested capacities will determine the renewable power productions in the future daily operations.

\subsection{Daily Operation}
Given the invested renewable capacities, each microgrid is responsible for the power scheduling in the microgrid as well as energy exchange with other interconnected microgrids. The operation horizon for the microgrid is one day, which is divided into $T=24$ equal time slots, denoted as $\mathcal{T}=\{1,...,T\}$. We assume that the operations of different days are independent, hence we will focus on the operation of one day in the rest of this subsection.\footnote{We assume that the users' power consumption and energy charging/discharging behaviors are repeated in a daily basis.}

In scenario $\omega$ and an operational horizon $\mathcal{T}$, microgrid $i$ determines the renewable power supply, main grid power procurement, and energy storage charging and discharging to meet its users' aggregate demand, which consists of both elastic and inelastic loads. In the following, we model the operational characteristics of microgrids, including supply model, energy storage model, demand model, and energy management of the interconnceted-microgrid system.

\subsubsection{Supply Model}
Microgrid has two sources for power supply: renewable power $\boldsymbol{g}_{i}^{\omega} = \{ g_{i}^{\omega,t},~\forall t \in \mathcal{T} \}$ and conventional power procurement $\boldsymbol{q}_{i}^{\omega} = \{ q_{i}^{\omega,t} ,~\forall t \in \mathcal{T} \}$. The power supplies satisfy the following constraints:
\begin{align}
 & 0 \leq g_{i}^{\omega,t} \leq z_{i} ( G_i^{s} \eta_{i}^{s,\omega,t} +G_i^{w} \eta_{i}^{w,\omega,t} ), \forall t\in\mathcal{T},~\forall i \in \mathcal{M}, \label{constraint-supply1} \\
 & 0 \leq q_{i}^{\omega,t} \leq Q_{i}^{ \max}, \forall t\in\mathcal{T},~\forall i \in \mathcal{M}, \label{constraint-supply2}
\end{align}
where $\boldsymbol{\eta}_{i}^{s,\omega} = \{ \eta_{i}^{s,\omega,t} ,\forall t \in \mathcal{T} \}$ and $\boldsymbol{\eta}_{i}^{w,\omega} = \{ \eta_{i}^{w,\omega,t} ,\forall t \in \mathcal{T} \}$ denote the solar and wind generations in microgrid $i$ under scenario $\omega$ per each unit of invested capacity. If $z_{i}=0$, microgrid $i$ does not install local renewable generation, and its local renewable supply is zero. If $z_{i}=1$, $G_i^{s} \eta_{i}^{s,\omega,t} +G_i^{w} \eta_{i}^{w,\omega,t}$ denotes the maximum available renewable power of microgrid $i$ in time slot $t$ under scenario $\omega$. Microgrid $i$ can curtail renewable power, and thus the actual renewable power supply $g_{i}^{\omega,t}$ in time slot $t$ can be less than the available renewable power $G_i^{s} \eta_{i}^{s,\omega,t} +G_i^{w} \eta_{i}^{w,\omega,t}$. For the main grid power supply, $q_{i}^{\omega,t}$ is bounded by $Q_{i}^{ \max}$, which denotes microgrid $i$’s maximum power procurement from the main grid. The microgrids are connected to the main grid through a point of common coupling (PCC), which could be distribution feeders, transformers or converters (for DC microgrids). The maximum power procurement of microgrid $i$ depends on the capacities of PCC and power bus within microgrid $i$. We assume that net metering is not allowed, which means that microgrids cannot sell power back to the main grid, \textit{i.e.,} $q_{i}^{\omega,t} \geq 0$.

\subsubsection{Energy Storage Model}
Energy storage (such as batteries) can smooth out the intermittent renewable power generation, and exploit time-varying operational costs for arbitrage. For microgrid $i$, we let $\boldsymbol{s}_{i}^{\omega} = \{ s_{i}^{\omega,t},~\forall t \in \mathcal{T} \}$, $\boldsymbol{r}_{i}^{\omega} = \{ r_{i}^{\omega,t},~\forall t \in \mathcal{T} \}$, and $\boldsymbol{d}_{i}^{\omega} = \{ d_{i}^{\omega,t},~\forall t \in \mathcal{T} \}$ denote the amount of electricity stored, charged, and discharged over the operational horizon $\mathcal{T}$ in scenario $\omega$, respectively.

First, the energy charging and discharging amounts are bounded, and satisfy the following constraints:
\begin{align}
 & 0 \leq r_{i}^{\omega,t} \leq r_{i}^{\max},~\forall t\in\mathcal{T},~\forall i \in \mathcal{M}, \label{constraint-storage1} \\
 & 0 \leq d_{i}^{\omega,t} \leq d_{i}^{\max},~\forall t\in\mathcal{T},~\forall i \in \mathcal{M}, \label{constraint-storage2}
\end{align}
where $r_{i}^{ \max} >0$ and $d_{i}^{ \max} >0$ denote the maximum charging and discharging limits, respectively.

Second, there are power losses when electricity is charged into and discharged from the battery. We denote $\eta_{i}^{r} \in ( 0,1 ]$ and $\eta_{i}^{d} \in ( 0,1 ]$ as the conversion efficiencies of charging and discharging. The battery's lifetime is heavily affected by the depth-of-discharge \cite{DOD}, and thus we introduce a maximum allowed depth-of-discharge $DoD_{i}$ to restrict the operation of battery. Specifically, the stored energy $s_{i}^{\omega,t}$ should be bounded between lower and upper bounds. We can set the upper bound $S_{i}^{\max}$ as the battery capacity $E_i$ in microgrid $i$. The lower bound can be set as $S_{i}^{\min} = E_i (1- DoD_{i})$, in which we can choose a low $DoD_{i}$ to reduce the impact of battery degradation. Therefore, we obtain microgrid $i$'s battery dynamics in time slot $t$ and scenario $\omega$ as
\begin{equation}
\begin{split} 
s_{i}^{\omega,t} = \min \left\{ \max \left\{ S_{i}^{\min}, s_{i}^{\omega,t-1} + \eta_{i}^{r} r_{i}^{\omega,t} - \frac{d_{i}^{\omega,t}} {\eta_{i}^{d}} \right\},S_{i}^{\max} \right\}, \\ 
\forall t \in \mathcal{T},~\forall i \in \mathcal{M}, \label{constraint-storage3}
\end{split}
\end{equation}
in which we further restrict the terminal energy level $s_{i}^{\omega,T}$ to be equal to the initial value $s_{i}^{\omega,0}$, such that the battery operation is independent across multiple operational horizons.

\subsubsection{Demand Model}
Let $\mathcal{N}_i$ denote the set of users in microgrid $i \in \mathcal{M}$. We classify the loads of each user $n \in \mathcal{N}_i$ into two categories: inelastic loads and elastic loads. The inelastic loads, such as refrigerator and illumination demands, cannot be easily shifted over time. We let $b_{i}^{t}$ denote the aggregate inelastic load of all the users in microgrid $i$ and time slot $t$, and denote $\boldsymbol{b}_{i} = \{ b_{i}^{t},~\forall t \in \mathcal{T} \}$. The elastic loads, such as HVAC (heating, ventilation and air conditioning) demand, electric vehicle and washing machine demands, can be flexibly scheduled over time. For a user $n \in \mathcal{N}_i$, we denote the elastic load as $\boldsymbol{x}_{n}^{\omega} = \{ x_{n}^{\omega,t} ,~\forall t \in \mathcal{T} \}$, where $x_{n}^{\omega,t}$ is user $n$'s elastic power consumption in slot $t$ under renewable generation scenario $\omega$.

The demand response program can only control the elastic loads, and should be subject to the following constraints:
\begin{align}
 & \sum_{t \in \mathcal{T}} x_{n}^{\omega,t} = L_{n},~\forall n \in \mathcal{N}_i,~\forall i \in \mathcal{M}, \label{constraint-demand1} \\
 & l_{n}^{t, \min} \leq x_{n}^{\omega,t} \leq l_{n}^{t, \max},\;\forall t\in\mathcal{T},~\forall n \in \mathcal{N}_i,~\forall i \in \mathcal{M}, \label{constraint-demand2}
\end{align}
where constraint (\ref{constraint-demand1}) corresponds to the prescribed total energy requirement $L_{n}$ in each day. Constraint (\ref{constraint-demand2}) provides a lower bound $l_{n}^{t, \min}$ and upper bound $l_{n}^{t, \max}$ for the power consumption of user $n$ in each time slot $t$.

\subsubsection{Operational Costs}
In each operational horizon (say a day) under renewable generation scenario $\omega$, microgrid $i$ coordinates its local power supply and demand by power supply scheduling, energy storage charging and discharging, and elastic load shifting through demand response program. Such power scheduling incurs an operational cost, including the costs of purchasing main grid power, energy storage operation, and demand response.

We assume that the cost of renewable power production is zero \cite{review1}, as renewable sources are free to utilize when the renewable generation facilities are installed and in operation. For the power purchased from the main grid, microgrid $i$ will be charged by a time-dependent unit price $p^t$ in time slot $t$, and thus the power supply cost of microgrid $i$ is written as $p^{t} q_{i}^{\omega,t}$ in time slot $t$ and scenario $\omega$.

Repeated charging and discharging cause degradation of the energy storage devices. We model the aging cost of energy storage as a function of charging and discharging amounts, and define the cost of energy storage operation \cite{storagecost} in microgrid $i$ as $\alpha_{i} \left(  r_{i}^{\omega,t} +  d_{i}^{\omega,t} \right)$ in time slot $t$ and scenario $\omega$, where $\alpha_{i}$ is the unit cost of energy storage charging and discharging in microgrid $i$.

Scheduling elastic load may affect user's comfort, as the scheduled power consumption deviates users' preferred power consumption. We let $\boldsymbol{y}_{n} = \{ y_{n}^{t} ,~\forall t \in \mathcal{T} \}$ denote the most preferred power consumption of user $n$, and define the discomfort cost \cite{discomfortcost} of user $n$ in time slot $t$ as $\beta_{n}  \left( x_{n}^{\omega,t}-y_{n}^{t} \right)^{2}$, where $\beta_{n}$ is used to indicate the sensitivity of user $n$ towards the power consumption deviation.

Therefore, we have the following operational cost of microgrid $i$ over the entire operational horizon in scenario $\omega$:
\begin{align*}
\begin{split}
 & C_{i}^{O} (\boldsymbol{q}_{i}^{\omega}, \boldsymbol{r}_{i}^{\omega}, \boldsymbol{d}_{i}^{\omega}, \boldsymbol{x}_{n}^{\omega}) \\
 = & \sum_{t\in\mathcal{T}} \left[ p^{t} q_{i}^{\omega,t} + \alpha_{i} \left(  r_{i}^{\omega,t} +  d_{i}^{\omega,t} \right) + \sum_{n \in \mathcal{N}_{i}} \beta_{n} \left( x_{n}^{\omega,t}-y_{n}^{t} \right)^{2} \right],
\end{split}
\end{align*}
which includes power procurement cost, battery operation cost and users' discomfort costs.

\section{Cooperative Planning of Renewable Generations}
As shown in Section II, microgrids in different locations have different renewable generation profiles and potentials. For example, when renewable generations in some locations are in deficit (relative to the demands at those locations), renewable generations at other locations could have significant surplus. Some locations have adequate renewable sources (\emph{e.g.}, high solar radiation or strong wind), while others do not. The fixed investment costs due to real estate are low in some suburban areas, but high in urban areas. All these factors will affect the economical planning and operation of renewable generations. Through cooperative planning and later utilization of renewable generation, microgrids can leverage the diversities of renewable generation profiles. Those microgrids with larger renewable generation capacities and excessive local renewable generations can supply power to other microgrids in short of local power supplies.

However, microgrids are often managed by local entities with local interests. They do not have incentives to over-invest in their local renewable generations and provide power supplies to other microgrids without proper incentives. To encourage cooperative planning among interconnected microgrids, we propose a cooperative planning and cost sharing scheme based on Nash bargaining solution \cite{nash}. Before presenting the cooperative planning model, we first present a non-cooperative benchmark problem in the following.

\subsection{Non-cooperative Benchmark}
We calculate the best performance (minimum overall cost) that each microgrid can achieve without cooperating with other microgrids. This corresponds to the outside option in the bargaining game, as each microgrid needs to decide whether to cooperate or not depending on whether the cooperation leads to a performance that is better than its corresponding outside option. In this noncooperative planning benchmark, microgrid $i$ balances its local power supply and demand, and minimizes its overall cost without interacting with other microgrids.

We assume that all the loads and renewable energy generations are connected to a common power bus within each microgrid, such that we can restrict the discussion on the balance between aggregate power supply and aggregate demand. The local power balance constraint for microgrid $i$ in time slot $t$ and scenario $\omega$ is
\begin{align}
g_{i}^{\omega,t} + q_{i}^{\omega,t} + d_{i}^{\omega,t} = r_{i}^{\omega,t} + b_{i}^{t} + \sum_{n \in \mathcal{N}_i} x_{n}^{\omega,t},~\forall t\in\mathcal{T}, \;\forall i\in\mathcal{M}. \label{constraint-balance1}
\end{align}

We denote the expected overall cost (\emph{i.e.}, investment plus operational costs) of microgrid $i$ over all possible scenarios $\omega \in \Omega$ as
\begin{equation*}
\begin{split} 
& C_{i}^{Overall} (z_i,G_i^{s},G_i^{w},\boldsymbol{q}_{i}^{\omega}, \boldsymbol{r}_{i}^{\omega}, \boldsymbol{d}_{i}^{\omega}, \boldsymbol{x}_{n}^{\omega}) \\
 \triangleq~ & C_i^{I}(z_i,G_i^{s},G_i^{w})
+ \theta \cdot \mathbb{E}_{\omega \in \Omega} C_{i}^{O} (\boldsymbol{q}_{i}^{\omega}, \boldsymbol{r}_{i}^{\omega}, \boldsymbol{d}_{i}^{\omega}, \boldsymbol{x}_{n}^{\omega}),
\end{split}
\end{equation*}
which consists of the initial investment cost $C_i^{I}(z_i,G_i^{s},G_i^{w})$ and the present value of the accumulative expected operational cost $\theta \cdot \mathbb{E}_{\omega \in \Omega} C_{i}^{O} (\boldsymbol{q}_{i}^{\omega}, \boldsymbol{r}_{i}^{\omega}, \boldsymbol{d}_{i}^{\omega}, \boldsymbol{x}_{n}^{\omega})$ in the entire planning horizon. The discounted coefficient for the operational cost $\theta$ is calculated by $\theta = \sum_{d=1}^{D} \frac{1}{(1 + R_d)^d}$, where $R_d$ is the daily discount rate. The expected daily operational cost can be calculated as
\begin{equation*}
\mathbb{E}_{\omega \in \Omega} C_{i}^{O} (\boldsymbol{q}_{i}^{\omega}, \boldsymbol{r}_{i}^{\omega}, \boldsymbol{d}_{i}^{\omega}, \boldsymbol{x}_{n}^{\omega}) 
=  \sum_{\omega \in \Omega} \pi_{\omega} C_{i}^{O} (\boldsymbol{q}_{i}^{\omega}, \boldsymbol{r}_{i}^{\omega}, \boldsymbol{d}_{i}^{\omega}, \boldsymbol{x}_{n}^{\omega}),
\end{equation*}
where the weight $\pi_{\omega}$ is the probability obtained in Section II.

We solve the expected overall cost minimization problem of microgrid $i$:
\begin{equation*}
\begin{aligned}
& \min 
&& C_{i}^{Overall} (z_i,G_i^{s},G_i^{w},\boldsymbol{q}_{i}^{\omega}, \boldsymbol{r}_{i}^{\omega}, \boldsymbol{d}_{i}^{\omega}, \boldsymbol{x}_{n}^{\omega}) \\
& \text{subject to}
&&  \eqref{constraint-supply1}-\eqref{constraint-balance1} , \\
& \text{variables:}
&& z_i, G_i^{s}, G_i^{w}, \boldsymbol{g}_{i}^{\omega}, \boldsymbol{q}_{i}^{\omega}, \boldsymbol{r}_{i}^{\omega}, \boldsymbol{d}_{i}^{\omega}, \boldsymbol{s}_{i}^{\omega}, \boldsymbol{x}_{n}^{\omega},
\end{aligned} 
\end{equation*}
and obtain the optimum denoted as $C_i^{NonCoop}$, which is the minimum expected overall cost that microgrid $i$ can achieve without cooperating with other. In Section V, we will compare the performances of the cooperative planning and this noncooperative benchmark.

\subsection{Cooperative Planning via Nash Bargaining}
Next we consider the cooperative renewable generation planning of interconnected microgrids. As shown in Section II, microgrids at different locations have different potentials and patterns of renewable power generations. Through cooperative planning and operation, interconnected-microgrids system can benefit from the diversity of renewable energy generations. However, each microgrid operator is a rational decision maker, and aims to optimize its own benefit (\emph{e.g.}, cost minimization). Therefore, we need to design a proper incentive mechanism to induce each microgrid to participate in the cooperative planning. We model the interactions among microgrids in the cooperative planning as a Nash bargaining game \cite{nash}.

First, in the planning period, interconnected microgrids cooperatively decide the renewable energy planning, and share the investment costs through bargaining. Let $\boldsymbol{v} = \{ v_{i},~\forall i \in \mathcal{M} \}$ be the cost sharing vector of all the microgrids. The summation of all the cost sharing should be equal to the total investment expense:
\begin{align}
\sum_{i \in \mathcal{M}} v_{i} = \sum_{i \in \mathcal{M}}  C_i^{I}(z_i,G_i^{s},G_i^{w}). \label{constraint-clearing}
\end{align}

Such an cost sharing scheme should not only covers the total investment expense, but also reflects the benefit gained by each microgrid in the operational period.

Second, when the renewable energy facilities are deployed, renewable power generations are dispatched to microgrids at different locations. Let $\boldsymbol{e}_{i}^{\omega} = \{ e_{i,j}^{\omega,t},~\forall t \in \mathcal{T},~\forall j \in \mathcal{M} \}$ denote the power supply vector for microgrid $i$, where $e_{i,j}^{\omega,t} \geq 0$ denotes the renewable power supply from microgrid $j$ to microgrid $i$. In practice, the power dispatch can be achieved by algorithm implemented in the control modules co-located with supply side (renewable generations) and demand side (microgrids). The total renewable power supply should be no greater than the available renewable power production, as shown in the following constraint:
\begin{equation}
 \sum_{j \in \mathcal{M}} e_{j,i}^{\omega,t} \leq z_{i} ( G_i^{s} \eta_{i}^{s,\omega,t} +G_i^{w} \eta_{i}^{w,\omega,t} ),~\forall t\in\mathcal{T}, \;\forall i \in\mathcal{M},
\label{constraint-renewablesupply}
\end{equation}
where $\sum_{j \in \mathcal{M}} e_{j,i}^{\omega,t}$ represents the total renewable power supply from microgrid $i$.

Note that the power distribution has loss, and we let $\eta_{i,j}$ denote the distribution efficiency between microgrid $j$ and microgrid $i$. For microgrid $i$, we have the new power balance constraint:
\begin{equation}
\begin{aligned}
&& \sum_{j \in \mathcal{M}} \eta_{i,j} e_{i,j}^{\omega,t} + q_{i}^{\omega,t} + d_{i}^{\omega,t}  = r_{i}^{\omega,t} + b_{i}^{t} + \sum_{n \in \mathcal{N}_i} x_{n}^{\omega,t},\\
&& \forall t\in\mathcal{T}, \;\forall i \in\mathcal{M},\label{constraint-balance}
\end{aligned}
\end{equation}
where the left-hand side and right-hand side of the equality constraint represent the net power supply and demand for microgrid $i$, respectively. The total renewable energy serving microgrid $i$ is represented by $\sum_{j \in \mathcal{M}} \eta_{i,j} e_{i,j}^{\omega,t}$. 

To guarantee that each microgrid is willing to participate in the cooperative planning, its overall cost should be less than that in the noncooperative benchmark. This leads to the following incentive constraint:
\begin{equation}
v_{i} + \theta \cdot \mathbb{E}_{\omega \in \Omega} C_{i}^{O} (\boldsymbol{q}_{i}^{\omega}, \boldsymbol{r}_{i}^{\omega}, \boldsymbol{d}_{i}^{\omega}, \boldsymbol{x}_{n}^{\omega})  \leq C_{i}^{NonCoop}, \\
~ \forall i \in \mathcal{M}, \label{constraint-incentive}
\end{equation}
where the overall cost of microgrid $i$ consists of its shared investment cost $v_i$ and the total expected operational cost.

We formulate the cooperative planning problem among $M$ interconnected microgrids as a Nash bargaining problem as
\begin{equation*}
\begin{split}
& \textbf{Cooperative Planning Problem (CPP)} \\
& \max~ \prod_{i\in\mathcal{M}} \left[ C_{i}^{NonCoop} - \left( v_{i} + \theta \cdot \mathbb{E}_{\omega \in \Omega} C_{i}^{O} (\boldsymbol{q}_{i}^{\omega}, \boldsymbol{r}_{i}^{\omega}, \boldsymbol{d}_{i}^{\omega}, \boldsymbol{x}_{n}^{\omega}) \right)  \right]\\
& \text{subject to} ~~ \eqref{constraint-supply2}-\eqref{constraint-demand2}  ~\text{and}~  \eqref{constraint-clearing}-\eqref{constraint-incentive} , \\
& \text{variables:} ~~\{ z_i, G_i^{s}, G_i^{w}, v_i, \boldsymbol{e}_{i}^{\omega}, \boldsymbol{q}_{i}^{\omega}, \boldsymbol{r}_{i}^{\omega}, \boldsymbol{d}_{i}^{\omega}, \boldsymbol{s}_{i}^{\omega}, \boldsymbol{x}_{n}^{\omega}, \forall i \in \mathcal{M} \}.
\end{split}
\end{equation*}

To solve Problem \textbf{CPP}, we have the following Theorem:
\begin{theorem}
We can solve Problem \textbf{CPP} in two steps:

Step 1: solve the joint investment and operation problem (\textbf{IOP}) of the system,
\begin{equation*}
\begin{aligned}
& \min 
&& \sum_{i\in\mathcal{M}} C_{i}^{Overall} (z_i,G_i^{s},G_i^{w},\boldsymbol{q}_{i}^{\omega}, \boldsymbol{r}_{i}^{\omega}, \boldsymbol{d}_{i}^{\omega}, \boldsymbol{x}_{n}^{\omega}) \\
& \text{subject to}
&&  \eqref{constraint-supply2}-\eqref{constraint-demand2},~\eqref{constraint-renewablesupply}~\text{and}~\eqref{constraint-balance}, \\
& \text{variables:}
&& \{ z_i, G_i^{s}, G_i^{w}, \boldsymbol{e}_{i}^{\omega}, \boldsymbol{q}_{i}^{\omega}, \boldsymbol{r}_{i}^{\omega}, \boldsymbol{d}_{i}^{\omega}, \boldsymbol{s}_{i}^{\omega}, \boldsymbol{x}_{n}^{\omega}, \forall i \in \mathcal{M} \},
\end{aligned} 
\end{equation*}
where we denote \( \displaystyle \{ z_i^{\star},G_i^{s,\star},G_i^{w,\star}, \forall i \in \mathcal{M} \} \) as the optimal planning, \( \displaystyle \{ \boldsymbol{e}_{i}^{\omega,\star}, \boldsymbol{q}_{i}^{\omega,\star}, \boldsymbol{r}_{i}^{\omega,\star}, \boldsymbol{d}_{i}^{\omega,\star}, \boldsymbol{s}_{i}^{\omega,\star}, \boldsymbol{x}_{n}^{\omega,\star}, \forall i \in \mathcal{M} \} \) as the optimal power schedule, and \( \displaystyle C_{i}^{Oper,\star} \triangleq \theta \cdot \mathbb{E}_{\omega \in \Omega} C_{i}^{O} (\boldsymbol{q}_{i}^{\omega,\star}, \boldsymbol{r}_{i}^{\omega,\star}, \boldsymbol{d}_{i}^{\omega,\star}, \boldsymbol{x}_{n}^{\omega,\star}) \) as the optimal minimum expected operational cost of microgrid $i$ over the entire planning horizon.

Step 2: given the optimal planning and operation decisions in Problem \textbf{IOP}, solve the cost sharing problem (\textbf{CSP}),
\begin{equation*}
\begin{aligned}
& \max 
&& \prod_{i\in\mathcal{M}} \left[ C_{i}^{NonCoop} - \left( C_{i}^{Oper,\star} + v_{i} \right)  \right] \\
& \text{subject to}
&& \eqref{constraint-clearing}  ~\text{and}~  \eqref{constraint-incentive}, \\
& \text{variables:}
&& \{ v_i, \forall i \in \mathcal{M} \}.
\end{aligned} 
\end{equation*}
\end{theorem}

\textbf{Theorem} 1 shows that the cooperative planning among microgrids through bargaining achieves the best overall performance for the distribution system. Problem \textbf{IOP} minimizes the overall cost of the microgrids-system, and solves the optimal investment in renewable generations and the optimal power scheduling of all microgrids. Given the optimal planning of renewable generations, we solve Problem \textbf{CSP} to derive the optimal cost sharing to incentivize cooperative planning. Problem \textbf{IOP} can be solved by mixed integer programming solver and Problem \textbf{CSP} can be solved by standard convex optimization techniques \cite{convex}. Note that the cost sharing scheme not only applies to the scenario where renewable generation facilities are planned at the same time, but also applies to the scenario where incremental capacity is built sequentially. The proof of \textbf{Theorem} 1 can be found in the appendix.

\section{Performance Evaluation}
In this section, we conduct numerical studies using realistic data of Hong Kong. We consider both noncooperative and cooperative cases, in which interconnected microgrids make renewable generation planning by themselves and cooperatively, respectively. We aim to study the benefit of cooperative planning, and to validate our proposed incentive mechanism for the interconnected-microgrid system.

\subsection{System Setup}
\begin{figure}[t]
	\centering
	\includegraphics[width=7.5cm]{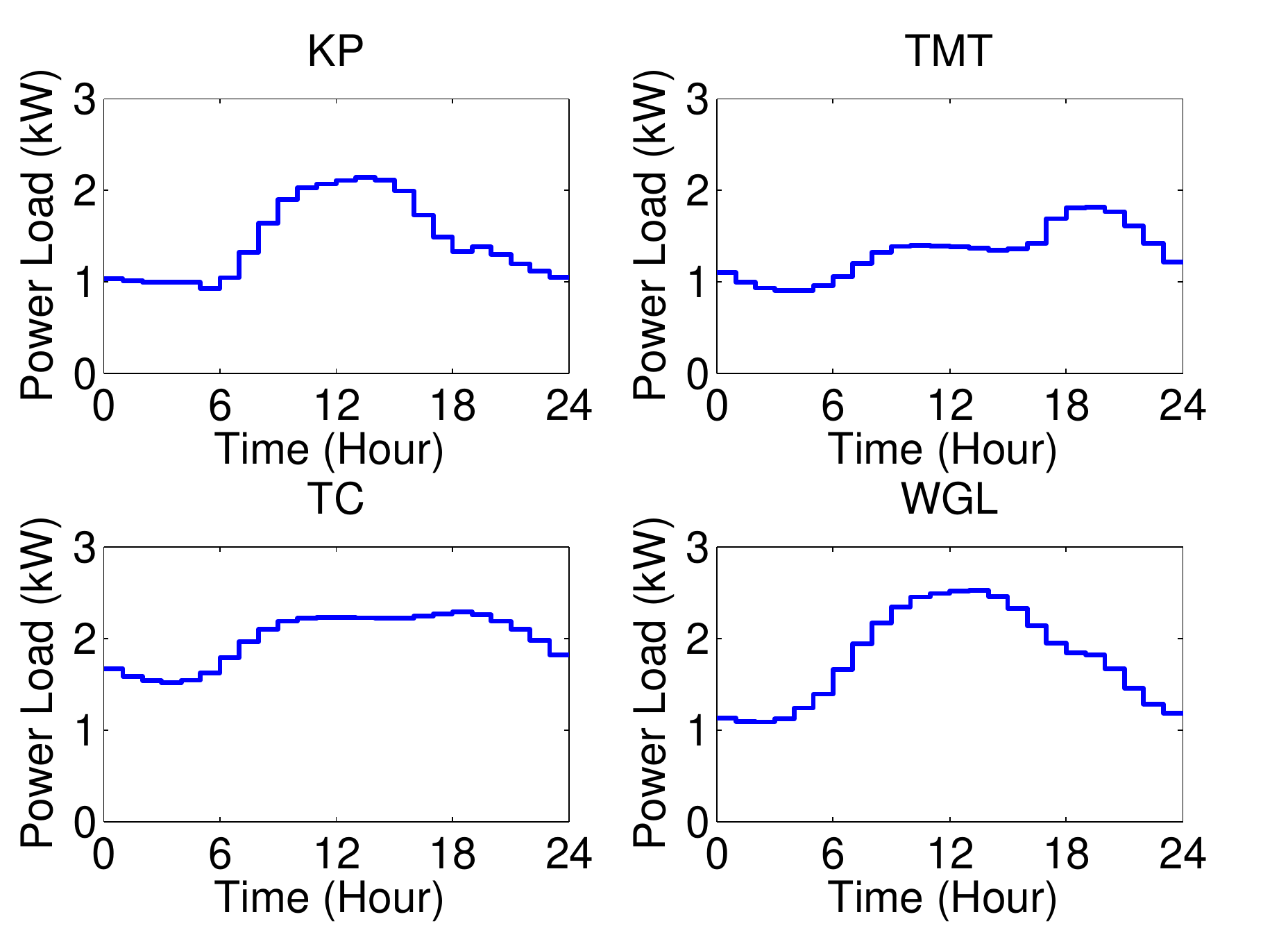}
	\caption{\label{fig-loads} Consumers' demands in four microgrids.} 
\end{figure}

We consider four interconnected microgrids, which are assumed to be located at KP, TMT, TC and WGL, respectively. Renewable generation scenarios at locations KP, TMT, TC and WGL illustrated in Fig. \ref{fig-scenario} are used to imitate the locational renewable generations in the four microgrids, respectively. Since our focus is on the renewable generation planning, we assume that each microgrid has equipped energy storage and demand response program. The users' loads are depicted in Fig. \ref{fig-loads}. We set the parameters of the solar power model and wind power model as in Table \ref{table-parameter}. Other main simulation parameters and users’ demand traces in each microgrid are summarized as follows: $H=20$, $R_{d} = 0.01$, fixed costs $F_{1} = 3.0 \times 10^{7}$, $F_{2} = 0.3 \times 10^{7}$, $F_{3} = 1.5 \times 10^{7}$, $F_{4} = 2.0 \times 10^{7}$, $\alpha_{i} = 0.2$, $\beta_{n} = 0.1$, $\eta_{i}^{r} = \eta_{i}^{d} = 0.98$, $Q_{i}^{ \max} = 5000$, $r_{i}^{ \max} = d_{i}^{ \max} = 0.2 \times S_{i}^{ \max} $, $G_i^{s,\max}=G_i^{w,\max}=5000$ $i, \in \{1,2,3,4 \}$.
\begin{table}[ht]
	\caption{Parameters of solar and wind power models}
	\label{table-parameter}
	\centering
	\begin{tabular}{c c}
		\hline \hline
		Model & Parameters \\
		\hline
		Solar power in (\ref{solarpower}) & $A_{m}=16$, $\eta_{m}=0.11$, $P_{f}=0.9$, $\eta_{c}=0.86$ \\
		Wind power in (\ref{windpower}) & $\rho=1.225$, $C_{p}=0.593$, $A=6.15$ \\
		\hline
	\end{tabular}
\end{table}

\subsection{Planning without Cooperation}
We first study the noncooperative benchmark, in which each microgrid itself decides whether or not to install its own renewable generations (solar and/or wind power), without interacting with other microgrids. Due to locational-diversity of renewable generation, microgrids have different conditions in terms of local solar and wind power profiles, and thus make very different decisions on whether to invest renewable generation or not. We calculate the optimal minimum overall cost for each microgrid, and derive the optimal strategy toward renewable generation planning.

For example, Table \ref{table-KP} and Table \ref{table-TMT} summarize the minimized overall costs of not deploying and deploying local renewable power facilities at KP and TMT, respectively. In terms of the renewable power profiles shown in Fig. \ref{fig-scenario}, both KP and TMT have low wind power potential. For KP, it is actually optimal not to install renewable energy but relies on main grid power. Table \ref{table-KP} shows that if KP chooses to deploy renewable power, then the best plan to invest 78.9M HKD; however, KP only gets 70.5M HKD in cost reduction in the operation, and the overall cost is 155.5M HKD, which is still greater than the overall cost without renewable. On the contrary, it is optimal for TMT to install renewable energy, as doing so will reduce the overall cost from 150.6M HKD to 137.1M HKD. 
\begin{table}[h]
\caption{Cost comparison at KP (in Million HKD)}
\label{table-KP}
\centering
\begin{tabular}{c| c| c}
\hline \hline
Cost &Without renewable &With renewable \\
\hline
Investment cost &0.0 & 78.9   \\ 	 	
Operational cost & 147.1 & 76.6   \\	 	
Overall cost & 147.1 & 155.5 \\
\hline
\end{tabular}
\end{table}

\begin{table}[h]
\caption{Cost comparison at TMT (in Million HKD)}
\label{table-TMT}
\centering
\begin{tabular}{c| c| c}
\hline \hline
Cost &Without renewable &With renewable \\
\hline
Investment cost &0.0 & 42.1   \\ 	 	
Operational cost & 150.6 & 95.0   \\	 	
Overall cost & 150.6 & 137.1 \\
\hline
\end{tabular}
\end{table}

Similarly, we show the minimized overall costs of not deploying and deploying local renewable power facilities at TC and WGL in Table \ref{table-TC} and Table \ref{table-WGL}, respectively. Both TC and WGL have high potential of renewable energy generation and a complementary relationship between wind power and solar power. It is optimal for both TC and WGL to install local renewable energy generations. Specifically, by investing 61.6M HKD in renewable energy, the operational cost at TC decreases dramatically from 303.8M HKD to 108.2M HKD, which implies that TC's demand can be mostly satisfied by its local renewable generation rather than the main grid. Table \ref{table-WGL} leads to a similar observation for WGL. 
\begin{table}[h]
\caption{Cost comparison at TC (in Million HKD)}
\label{table-TC}
\centering
\begin{tabular}{c| c| c}
\hline \hline
Cost &Without renewable &With renewable \\
\hline
Investment cost &0.0 & 61.6   \\ 	 	
Operational cost & 303.8 & 46.6   \\	 	
Overall cost & 303.8 & 108.2 \\
\hline
\end{tabular}
\end{table}

\begin{table}[h]
\caption{Cost comparison at WGL (in Million HKD)}
\label{table-WGL}
\centering
\begin{tabular}{c| c| c}
\hline \hline
Cost &Without renewable &With renewable \\
\hline
Investment cost &0.0 & 59.0   \\ 	 	
Operational cost & 251.0 & 42.0   \\	 	
Overall cost & 251.0 & 101.0 \\
\hline
\end{tabular}
\end{table}

\subsection{Cooperative Planning}
From the noncooperative benchmark analysis above, we can see that different microgrids exhibit various differences in renewable generation planning behaviors. Next we study the cooperative planning, in which microgrids coordinate with each other to determine the social optimal planning and fair cost sharing.

\begin{figure}[h]
	\centering
    \includegraphics[width=7.0cm]{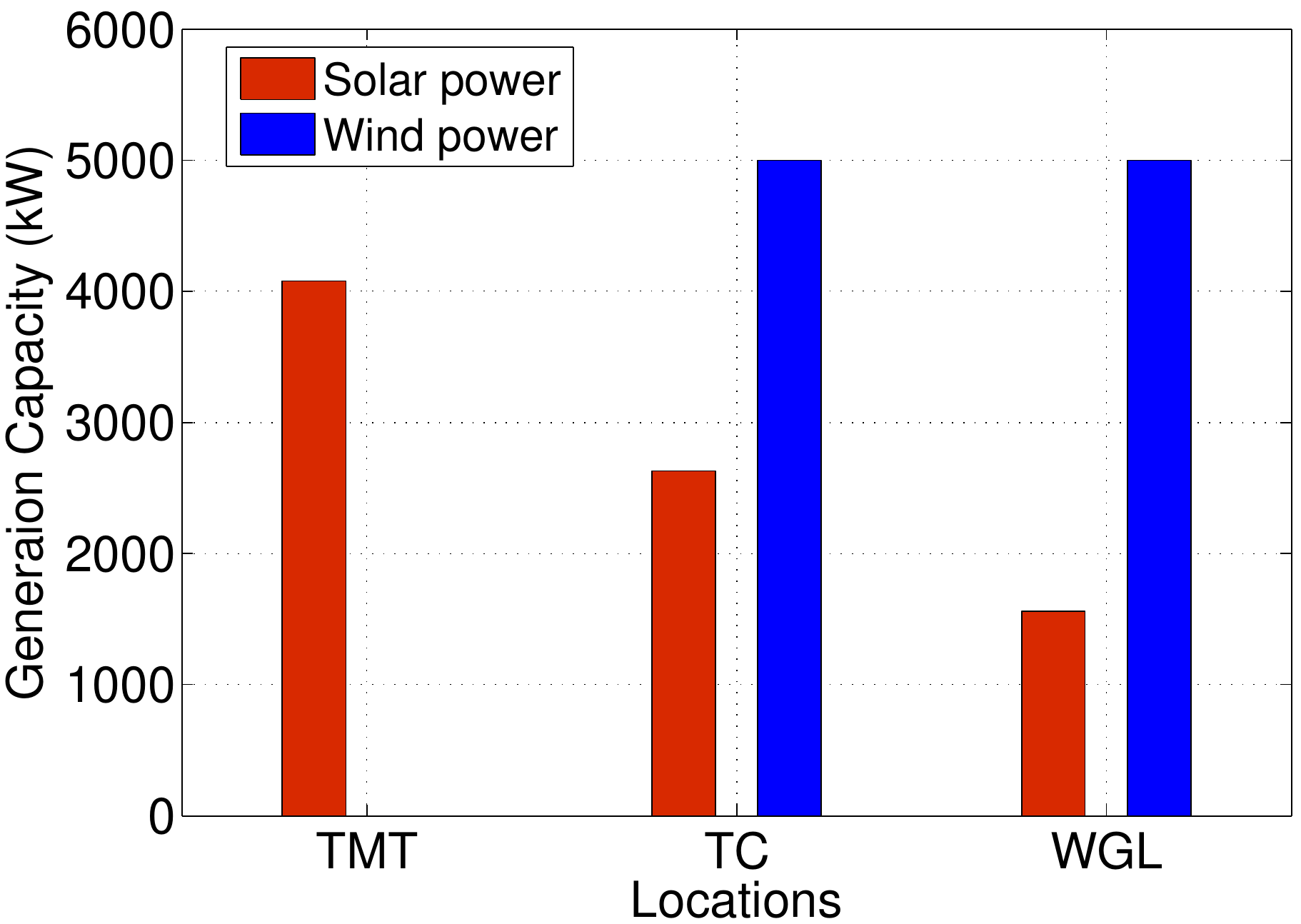}
    \caption{\label{fig-capa} Optimal planning of renewable energy.}
\end{figure}

In Fig. \ref{fig-capa}, we plot the optimal renewable energy planning (including solar power and wind power) for the interconnected-microgrids system. The optimal cooperative planning does not install any renewable energy at KP, as the fixed investment cost at KP is high, and meanwhile other three locations can provide adequate renewable energy for the entire system. At TC and WGL, both solar power and wind power are invested, and wind power has a larger invested capacity than solar power. This is because wind power at TC and WGL has a higher average power output than solar power. On the contrary, at TMT, only solar power is invested, because the solar power produces more compared to the wind power at TMT. Through cooperative planning, microgrids are able to take full advantage of the diverse renewable resources and improve the social welfare. The overall cost of the system (investment and operational costs) is reduce by 35.9\% compared to the overall cost of all the microgrids under noncooperative planning.

\begin{figure}[h]
	\centering
	\includegraphics[width=7.0cm]{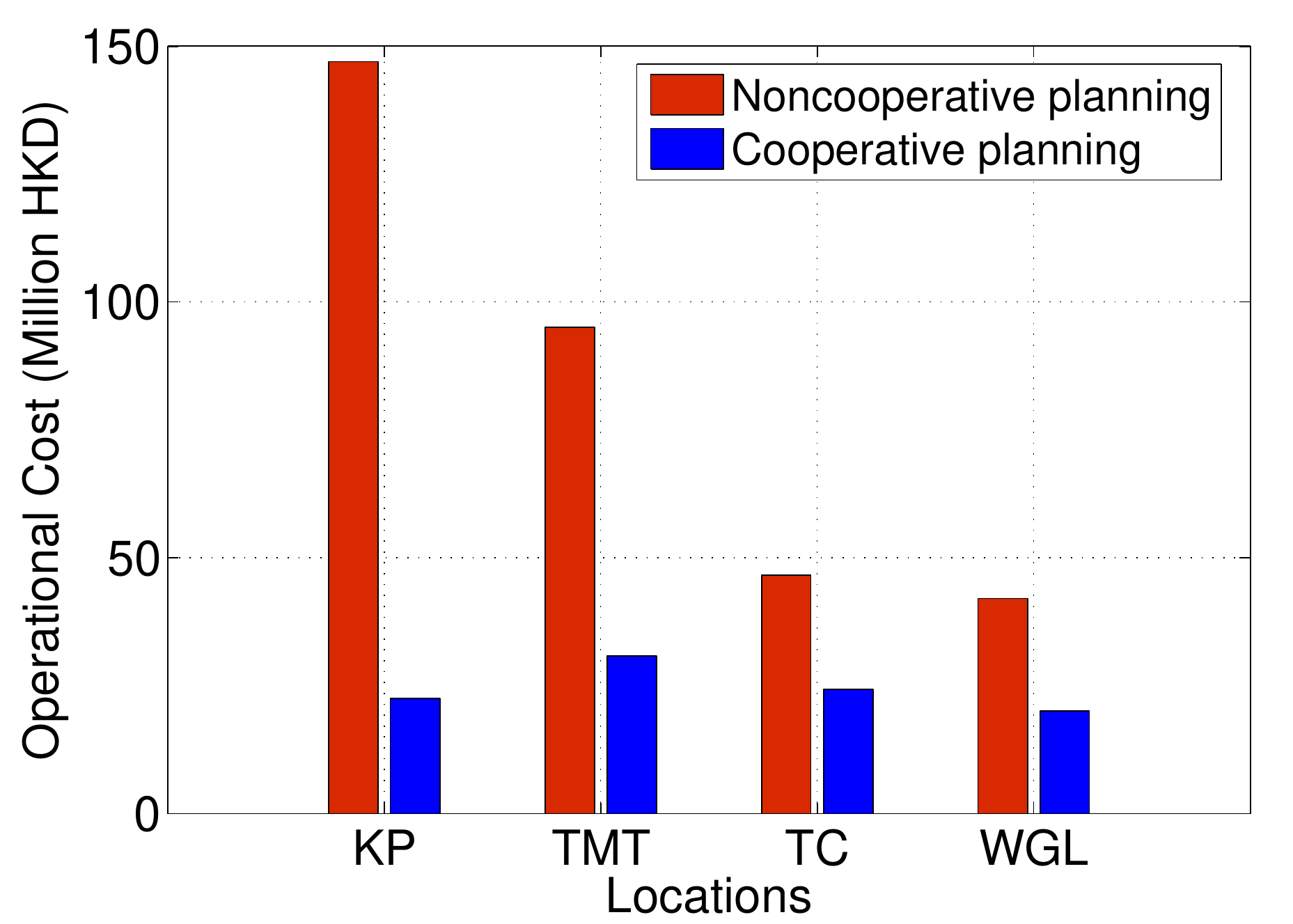}
	\caption{\label{fig-operationalcost} Comparison of operational cost.}
\end{figure}

\begin{figure}[h]
	\centering
	\includegraphics[width=7.0cm]{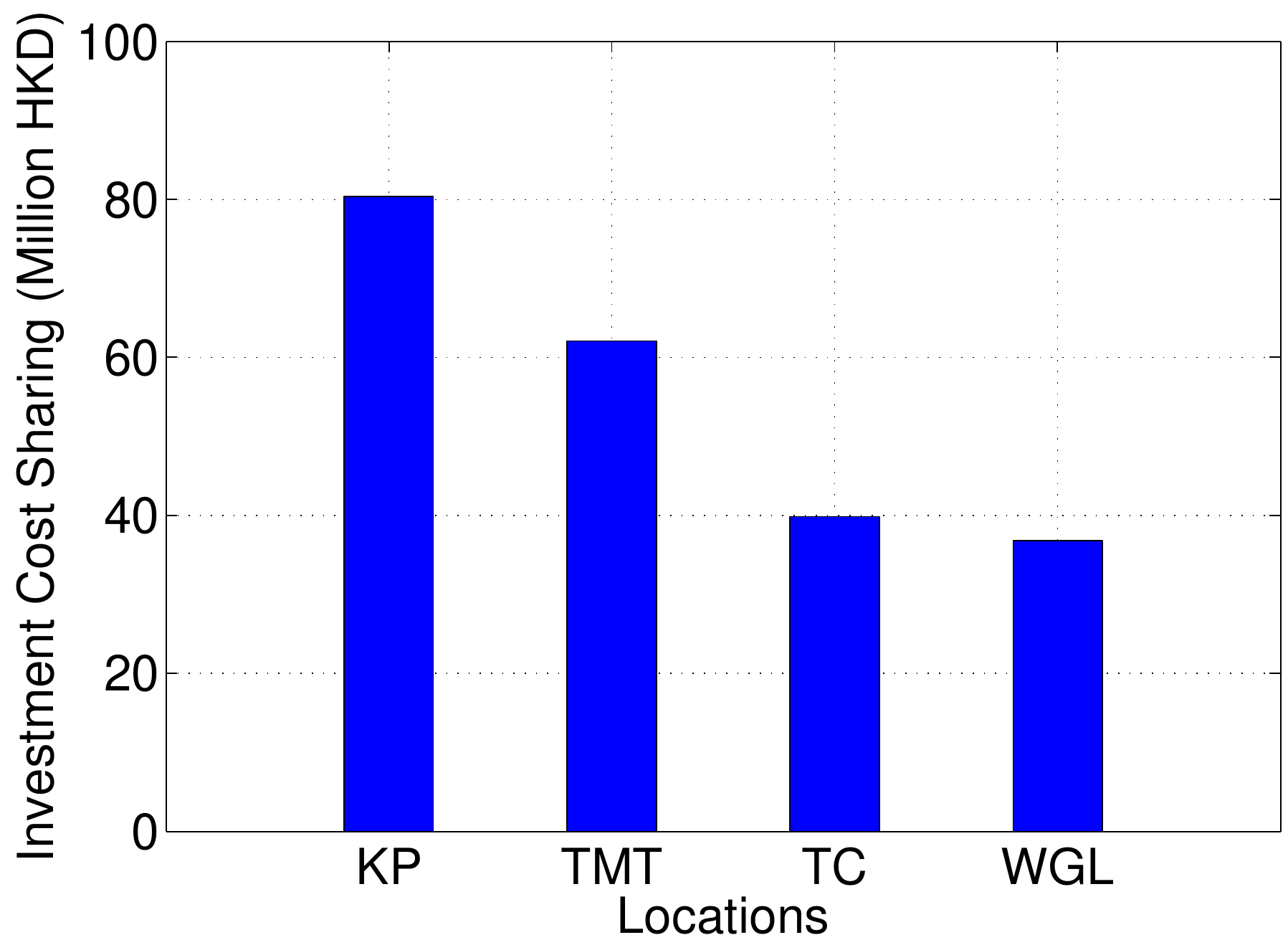}
	\caption{\label{fig-costsharing} Optimal cost sharing.} 
\end{figure}

In Fig. \ref{fig-operationalcost}, we compare the operational costs under noncooperative and cooperative plannings. The cooperative planning significantly reduces the operational cost of each microgrid, especially those who do not have high local potential of renewable energy generation potential (\emph{e.g.}, KP and TMT). For example, it is not economical for KP to deploy local renewable energy in the noncooperative case. Instead, KP has a strong incentive to participate in the cooperative planning and pay for others in order to get renewable energy supply (also see Fig. \ref{fig-costsharing} later on). As a result, KP reduces its operational cost by more than 4/5 through cooperation. For TC and WGL, they are able to benefit significantly from high local renewable energy generation even in the noncooperative case, and hence the additional gains from cooperation are small. 

In Fig. \ref{fig-costsharing}, we plot the optimal cost sharing derived from Nash bargaining solution. Cost sharing relies on the operational cost reduction between non-cooperative and cooperative scenarios. Fig. \ref{fig-operationalcost} shows that KP gains significant cost reduction (from 147.1M to 22.5M HKD) through cooperating with other microgrids. Similarly, TMT also gains significant cost reduction (from 150.6M to 30.8M HKD) through cooperation. Therefore, KP and TMT share the largest portions of the total investment cost, as they benefit most from the cooperation (as discussed in Fig. \ref{fig-operationalcost}). Relatively speaking, TC and WGL benefit less from the cooperation, and hence their shares of the investment cost are smaller than KP and TMT. The cost sharing is fair as those who benefit more need to share more investment cost. The overall costs (shared investment cost plus operational cost) of all microgrids are reduced by 30\%-44\%, such that all the microgrids are better off in the cooperative planning. This demonstrates that our proposed cost sharing scheme provides incentives to all the interconnected microgrids toward cooperative planning. 

\section{Conclusion}
We proposed a theoretical framework to study the cooperative planning of renewable generations in a distribution network, considering variable nature of renewable energy generations, self-interested behaviors of microgrids, and both long-term investment and short-term operation of the system. We analyzed the renewable energy generations using realistic meteorological data of Hong Kong. We designed an incentive mechanism, which encourages cooperation among interconnected microgrids towards a socially optimal planning, and splits the total investment cost in a fair manner. Simulation studies based on realistic data characterized the optimal investment decisions, and demonstrated the economic benefit (with 35.9\% overall cost reduction) of the cooperative planning method. In our future work, we are interested in the interactions not only among microgrids but also between the microgrids-group and the main grid.

\appendix
\subsection{Proof of Theorem 1}
First, we divide the decision variables of microgrid $i$ into the joint planning and operational decisions $\{ z_i, G_i^{s}, G_i^{w}, \boldsymbol{e}_{i}^{\omega}, \boldsymbol{q}_{i}^{\omega}, \boldsymbol{r}_{i}^{\omega}, \boldsymbol{d}_{i}^{\omega}, \boldsymbol{s}_{i}^{\omega}, \boldsymbol{x}_{n}^{\omega},~\forall n \in \mathcal{N}_i ,~\forall \omega \in \Omega \}$ and planning cost sharing decision $v_i$.

We can characterize the optimal solution of Problem \textbf{CPP} as follows. Given the optimal joint planning and operational decisions \( \displaystyle \{ z_i^{\star},G_i^{s,\star},G_i^{w,\star},  \boldsymbol{e}_{i}^{\omega,\star}, \boldsymbol{q}_{i}^{\omega,\star}, \boldsymbol{r}_{i}^{\omega,\star}, \boldsymbol{d}_{i}^{\omega,\star}, \boldsymbol{s}_{i}^{\omega,\star}, \boldsymbol{x}_{n}^{\omega,\star}, ~\forall n \in \mathcal{N}_i ,~\forall \omega \in \Omega, ~\forall i \in \mathcal{M} \} \), we can solve the optimal cost sharing decisions $\{ v_i^{\star},~\forall i \in \mathcal{M} \}$ through
\begin{equation}
\begin{aligned}
& \max 
&& \prod_{i\in\mathcal{M}} \left[ C_{i}^{NonCoop} - \left( v_{i} + C_{i}^{Oper,\star} \right)  \right] \\
& \text{subject to}
&& \eqref{constraint-clearing}  ~\text{and}~  \eqref{constraint-incentive}, \\
& \text{variables:}
&& \{ v_i,~\forall i \in \mathcal{M} \},
\label{problem-costsharing}
\end{aligned} 
\end{equation}
where the minimum expected operational cost of microgrid $i$ is denoted by \( \displaystyle C_{i}^{Oper,\star} \triangleq \theta \cdot \mathbb{E}_{\omega \in \Omega} C_{i}^{O} (\boldsymbol{q}_{i}^{\omega,\star}, \boldsymbol{r}_{i}^{\omega,\star}, \boldsymbol{d}_{i}^{\omega,\star}, \boldsymbol{x}_{n}^{\omega,\star}) \).

Solving \eqref{problem-costsharing}, we obtain the following relation between the optimal joint planning and operational decisions and the optimal cost sharing decisions $v_{i}^{\star}$:
\begin{equation}
\begin{split}
   & C_{i}^{NonCoop} - \left( v_{i}^{\star} + C_{i}^{Oper,\star} \right) \\
 = & \frac{ \sum_{i\in\mathcal{M}} \left[ C_{i}^{NonCoop} - \left( C_i^{I}(z_i^{\star},G_i^{s,\star},G_i^{w,\star}) + C_{i}^{Oper,\star} \right) \right] } 
{M}. \label{relation} 
\end{split}
\end{equation}

Substituting (\ref{relation}) into Problem \textbf{CPP} yields the optimal objective of the cooperative planning problem:
\begin{equation}
\prod_{i\in\mathcal{M}} \left[  \frac{\sum_{i\in\mathcal{M}} \left( C_{i}^{NonCoop} -  C_{i}^{Overall,\star} \right)}  {M} \right]^{M}, \label{optimalobjective}
\end{equation}
where the optimal overall cost of microgrid $i$ is denoted by \( \displaystyle C_{i}^{Overall,\star} \triangleq C_i^{I}(z_i^{\star},G_i^{s,\star},G_i^{w,\star}) + C_{i}^{Oper,\star} \).

From (\ref{optimalobjective}), we conclude that Problem \textbf{CPP} maximizes the social benefit of all the microgrids, \emph{i.e.}, \( \displaystyle \sum_{i\in\mathcal{M}} \left( C_{i}^{NonCoop} -  C_{i}^{Overall,\star} \right)\), through cooperative planning of renewable generations. Since $C_{i}^{NonCoop}$ is given, we prove that Problem \textbf{CPP} minimizes the social overall cost of all the microgrids, \emph{i.e.}, \( \displaystyle \sum_{i\in\mathcal{M}} C_{i}^{Overall,\star} \).

Therefore, we can decompose the original cooperative planning problem \textbf{CPP} into two consecutive problems. First, we minimize the social cost of the microgrids-system by solving joint investment and operation problem (\textbf{IOP}),
\begin{equation*}
\begin{aligned}
& \min 
&& \sum_{i\in\mathcal{M}} C_{i}^{Overall} (z_i,G_i^{s},G_i^{w},\boldsymbol{q}_{i}^{\omega}, \boldsymbol{r}_{i}^{\omega}, \boldsymbol{d}_{i}^{\omega}, \boldsymbol{x}_{n}^{\omega}) \\
& \text{subject to}
&& \eqref{constraint-supply2}-\eqref{constraint-demand2},~\eqref{constraint-renewablesupply}~\text{and}~\eqref{constraint-balance}, \\
& \text{variables:}
&& \{ z_i, G_i^{s}, G_i^{w}, \boldsymbol{e}_{i}^{\omega}, \boldsymbol{q}_{i}^{\omega}, \boldsymbol{r}_{i}^{\omega}, \boldsymbol{d}_{i}^{\omega}, \boldsymbol{s}_{i}^{\omega}, \boldsymbol{x}_{n}^{\omega}, \forall i \in \mathcal{M} \},
\end{aligned} 
\end{equation*}
and then we solve the cost sharing problem (\textbf{CSP}), given the optimal operational cost of each microgrid \( \displaystyle C_{i}^{Oper,\star} \) and optimal total planning cost \( \displaystyle \sum_{i\in\mathcal{M}}C_i^{I}(z_i^{\star},G_i^{s,\star},G_i^{w,\star})\),
\begin{equation*}
\begin{aligned}
& \max 
&& \prod_{i\in\mathcal{M}} \left[ C_{i}^{NonCoop} - \left( C_{i}^{Oper,\star} + v_{i} \right)  \right] \\
& \text{subject to}
&& \eqref{constraint-clearing}  ~\text{and}~  \eqref{constraint-incentive}, \\
& \text{variables:}
&& \{ v_i, \forall i \in \mathcal{M} \}.
\end{aligned} 
\end{equation*}

\end{document}